\documentclass[twocolumn, trackchanges, twocolappendix]{aastex631}
\usepackage{graphicx}
\usepackage{float}
\usepackage{enumitem}
\usepackage{natbib}
\usepackage{multirow}
\usepackage{CJK}

\shortauthors{L. Wang et al.}

\defcitealias{Li2018}{Li18}

\graphicspath{{./}{figures/}}
\begin{document}
\begin{CJK*}{UTF8}{gbsn}

\title{Unveiling Bifurcated Blue Straggler Sequences in NGC\,2173: Insights from Binary Evolution}
\email{wangli79@mail2.sysu.edu.cn;\\
dengkai@ynao.ac.cn}

\author[0000-0003-3471-9489]{Li Wang (王莉)}
\affiliation{School of Physics and Astronomy, Sun Yat-sen University, Daxue Road, Zhuhai, 519082, People's Republic of China}
\affiliation{CSST Science Center for the Guangdong--Hong Kong--Macau Greater Bay Area, Zhuhai, 519082, People's Republic of China}

\author[0000-0003-4265-7783]{Dengkai Jiang (姜登凯)}
\affiliation{Yunnan Observatories, Chinese Academy of Sciences, 396 Yangfangwang, Guandu District, Kunming 650216, People's Republic of China}
\affiliation{Center for Astronomical Mega-Science, Chinese Academy of Sciences, 20A Datun Road, Chaoyang District, Beijing 100012, People's Republic of China}
\affiliation{Key Laboratory for the Structure and Evolution of Celestial Objects, Chinese Academy of Sciences, Kunming 650011, People's Republic of China}
\affiliation{University of Chinese Academy of Sciences, Beijing 100049, People's Republic of China}

\author[0000-0002-3084-5157]{Chengyuan Li (李程远)}
\affiliation{School of Physics and Astronomy, Sun Yat-sen University, Daxue Road, Zhuhai, 519082, People's Republic of China}
\affiliation{CSST Science Center for the Guangdong--Hong Kong--Macau Greater Bay Area, Zhuhai, 519082, People's Republic of China}

\author[0000-0001-9073-9914]{Licai Deng (邓李才)}
\affiliation{Key Laboratory for Optical Astronomy, National Astronomical Observatories, Chinese Academy of Sciences, Beijing 100101, People's Republic of China}
\affiliation{School of Astronomy and Space Science, University of Chinese Academy of Sciences, Beijing 100049, People's Republic of China}
\affiliation{Department of Astronomy, School of Physics and Astronomy, China West Normal University, Nanchong 637002, People's Republic of China}

\author[0000-0001-7506-930X]{Antonino P. Milone}
\affil{Dipartimento di Fisica e Astronomia ``Galileo Galilei'', Univ. di Padova, Vicolo dell'Osservatorio 3, Padova, IT-35122, Italy}
\affil{Istituto Nazionale di Astrofisica - Osservatorio Astronomico di Padova, Vicolo dell'Osservatorio 5, Padova, IT-35122, Italy}

\author[0000-0001-8713-0366]{Long Wang (王龙)}
\affiliation{School of Physics and Astronomy, Sun Yat-sen University, Daxue Road, Zhuhai, 519082, People's Republic of China}
\affiliation{CSST Science Center for the Guangdong--Hong Kong--Macau Greater Bay Area, Zhuhai, 519082, People's Republic of China}

\begin{abstract}
Identifying bifurcated blue straggler (BS) sequences in color-magnitude diagrams (CMDs) of star clusters has long been regarded as a powerful diagnostic for distinguishing different BS formation mechanisms. While such bifurcations are typically associated with core-collapsed clusters, their detection in dynamically young clusters raises new questions about their origins. In this study, using high-precision proper motion data derived from {\sl Hubble Space Telescope} multi-epoch observations, we confirm the existence of two distinct BS sequences in the Large Magellanic Cloud cluster NGC\,2173 ($\sim$ 1.58 Gyr): a well-defined, narrow blue sequence and a sparser red sequence. The extended core region excludes collisional formation as a viable channel for BS formation. Our binary evolution models suggest that non-conservative mass transfer (MT) is essential for shaping the distribution of BS binaries in the CMD. The red sequence primarily comprises BS binaries formed through conservative ongoing MT. In contrast, the blue sequence BS binaries are formed through non-conservative post-MT processes, all involving white dwarf (WD) companions. These BS+WD binary systems may subsequently undergo a second MT phase, leading to the formation of double WD systems. 

\end{abstract}


\section{Introduction} \label{sec:introduction}

Blue stragglers (BS) are anomalous stars that are typically brighter and bluer than the main-sequence turnoff (MSTO) in the color-magnitude diagrams (CMDs) of the host star clusters. Their position suggests they are more massive than their MSTO counterparts, contradicting predictions of standard single-star evolution. Initially discovered in the globular cluster (GC) M\,3 by \citet{Sandage1953}, BSs have since been widely observed in various stellar systems, including all GCs \citep[e.g.,][]{Piotto2004, Leigh2007, Ferraro2018}, open clusters \citep[OCs; e.g.,][]{Mathieu2009}, dwarf galaxies \citep{Mapelli2007}, and the Galactic field \citep[e.g.,][]{Clarkson2011, Schirbel2015, Ekanayake2018}.

The formation mechanisms of BS have been a focus of extensive research, revealing two primary channels: binary interactions, which include mass transfer (MT) and coalescence within binary systems \citep{McCrea1964, Andronov2006}, and direct stellar collisions \citep{Hills1976}. It has been suggested that binary evolution predominates in the formation of BSs, regardless of the dynamical state of the host cluster \citep[e.g.,][]{Knigge2009, Sun2018}. The additional mass of BSs generally involves the interactions of two or more stars. Triple and binary-binary systems appear to play a crucial role in the formation of merger BSs \citep[e.g.][]{Fregeau2004, Perets2009}. Moreover, a large fraction of BSs may be generated by stellar mergers that occurred during the early cluster evolution, through tidal interactions with their circumstellar material, as suggested by \citet{Wang2022}. BSs would exhibit distinct characteristics in multiplicity \citep[e.g.,][]{Ferraro2009, Gosnell2014, Bodensteiner2021, Panthi2023}, rotation \citep[e.g.,][]{Bodensteiner2023, Ferraro2023, Billi2023}, chemical composition \citep[e.g.,][]{Ferraro2006, Lovisi2010, Milliman2015, Schneider2019}, and magnetic fields \citep{Schneider2019}, depending on their formation mechanisms. 

These formation channels are not mutually exclusive, however, the relative contributions of direct stellar collisions and binary evolution to the formation of BSs in clusters remain an open question. One of the most compelling discoveries in recent BS studies to constrain their formation is the presence of two distinct sequences of BSs in the CMDs of certain old GCs, often referred to as the blue sequence and red sequence \citep{Ferraro2009, Dalessandro2013, Simunovic2014, Beccari2019}. \citet{Ferraro2009} proposed that the coexistence of these two sequences arises from the simultaneous enhancement of two distinct formation mechanisms during core collapse, with each sequence corresponding to a different mechanism. In this scenario, the narrow blue sequence is attributed to BSs formed through direct stellar collisions during the core collapse, while the broad red sequence represents BSs originating from binary evolution. 

Whether the double BS sequences represent a feature of the core collapse event remains highly debated. It is worth noting that three W UMa contact binaries, commonly associated with binary evolution \citep{Rucinski2000, Jiang2013}, were detected in both two BS sequences of M\,30 \citep{Ferraro2009}. NGC\,1261 (11.5 Gyr), as one of the GCs exhibiting two BS sequences, does not show the classical features of the core collapse \citep{Simunovic2014, Raso2020}. Unlike old GCs, where the formation of BSs is heavily influenced by the current dynamical state of the cluster, BSs in OCs are primarily the products of binary evolution \citep[e.g.,][]{Sollima2008, Cordoni2023, Mohandasan2024, Rain2024}. However, the characteristics of the two sequences have been progressively identified in Galactic OC NGC\,7789 \citep[1.6 Gyr;][]{Rao2022}, Berkeley\,17 \citep[8.5--10 Gyr;][]{Rao2023}, and Melotte\,66 \citep[4.7 Gyr;][]{Rain2024}. Several theoretical studies have demonstrated that binary evolution alone can explain both sequences without invoking core collapse. \citet{Jiang2017} and \citet{Jiang2022} showed that binary evolution can contribute to the BSs in different regions of the CMDs of old GCs with various ages ($\geq$ 10 Gyr): blue-sequence BSs can originate from post-MT systems, where a rejuvenated BS orbits a white dwarf (WD), while red-sequence BSs often involve ongoing MT in close binaries. These findings challenge the traditional view that core collapse is necessary to explain the bifurcation, instead highlighting the significant role of binary evolution across different cluster environments.

\citet[][hereafter \citetalias{Li2018}]{Li2018} first identified the visually discernible bifurcated BS populations in the young massive cluster NGC\,2173 (1.58 Gyr) in the Large Magellanic Cloud (LMC), with no evidence of a core-collapse event. Due to the lack of proper motion (PM) information for stars in the cluster NGC\,2173 region at that time, \citet{Dalessandro2019a, Dalessandro2019b} strongly argued the bifurcated BS pattern in NGC\,2173 detected by \citetalias[]{Li2018} is most likely an artifact of field contamination. Thanks to multi-epoch observations with the {\sl Hubble Space Telescope} ({\sl HST}), it has become possible to differentiate member stars from the field stars within MC clusters based on their PMs, thus allowing accurate cluster membership determination. In this study, we revisit the young massive cluster NGC\,2173 and, through stellar kinematics, confirm the findings of \citetalias[][]{Li2018} that the bifurcated BS sequences they identified consist of genuine cluster members. We also investigate the origin of the definite bifurcated BSs in this young massive star cluster in detail, focusing on their formation through the binary MT scenario at the age of NGC\,2173. 

This article is structured as follows: Section \ref{sec: observations} describes the observations and data reduction procedures. Section \ref{sec: results} presents the main observational results. The binary population synthesis and corresponding outcomes are detailed in Section \ref{sec: syntheticpop}. Finally, the discussion and summary are presented in Section \ref{sec: discussion} and \ref{sec: summary}, respectively.

\section{Observations} \label{sec: observations}

 \begin{figure*}[ht!]
 \centering
 \epsscale{1.15}
 \plotone{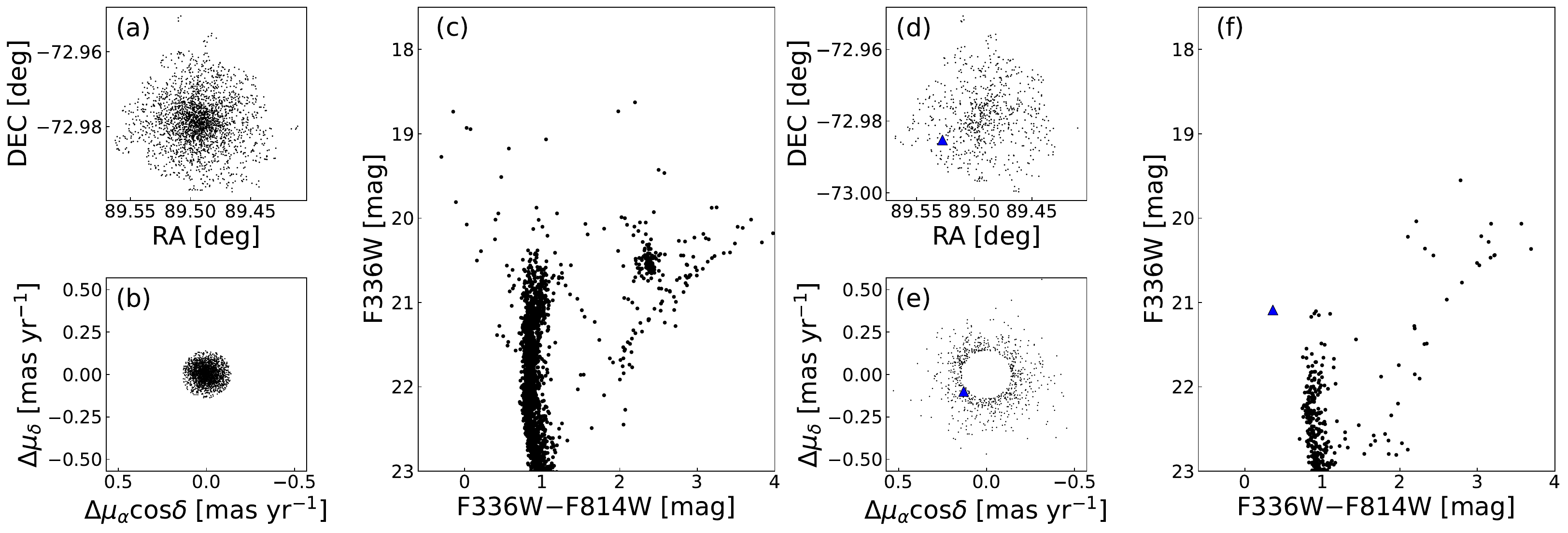}
 \caption{(a) Spatial distribution and (b) Relative PM diagram of member stars towards NGC\,2173. (c) Field-star decontaminated CMD of NGC\,2173. (d-f) As panels (a-c), but for field stars, are then subtracted from the raw stellar catalog. The blue triangle represents the only BS candidate excluded based on its PM.}
 \label{f1}
 \end{figure*}

\subsection{Membership Determination}

\begin{table*}
\caption{Detail information of the {\sl HST} datasets used in this work}
\label{T1}
\centering
\begin{tabular}{ccccc}
\hline\hline
Instrument & Filters & Exposure Time & Start/Stop Time & Proposal ID \\
\hline
  UVIS/WFC3 & F336W & 2$\times$700 s + 800 s = 2200 s & 2011-10-09 & 12257\\
  UVIS/WFC3 & F814W & 30 s + 2$\times$700 s + 550 s = 1980 s & 2011-10-09 & 12257\\
  UVIS/WFC3 & F814W & 2$\times$40 s + 4$\times$454 s + 455 s = 2351 s & 2022-06-28 & 16748\\
\hline
\end{tabular}
\end{table*}

With the increasing accumulation of observational data from the {\sl HST}, significant progress has been made in distinguishing member stars from field stars within Magellanic Cloud (MC) clusters. In this study, we utilize the NGC\,2173 stellar catalog from \citet{Milone2023}, which provides both astrometry and photometry for each star. This catalog is based on high-precision, multi-epoch {\sl HST} observations taken over a time interval of $\sim$ 10 yr in the F336W and F814W bands of the Ultraviolet and Visual Channel of the Wide Field Camera 3 (UVIS/WFC3). Detailed information about the {\sl HST} datasets and the methods used to estimate PMs are outlined in our Table \ref{T1} and in \citet{Milone2023}, respectively. The {\sl HST} data presented in this article were obtained from the Mikulski Archive for Space Telescopes (MAST) at the Space Telescope Science Institute. The specific observations analyzed can be accessed via \dataset[doi: 10.17909/zwfd-q641]{https://doi.org/10.17909/zwfd-q641}.

\begin{figure}[ht!]
\centering
\epsscale{1.15}
 \plotone{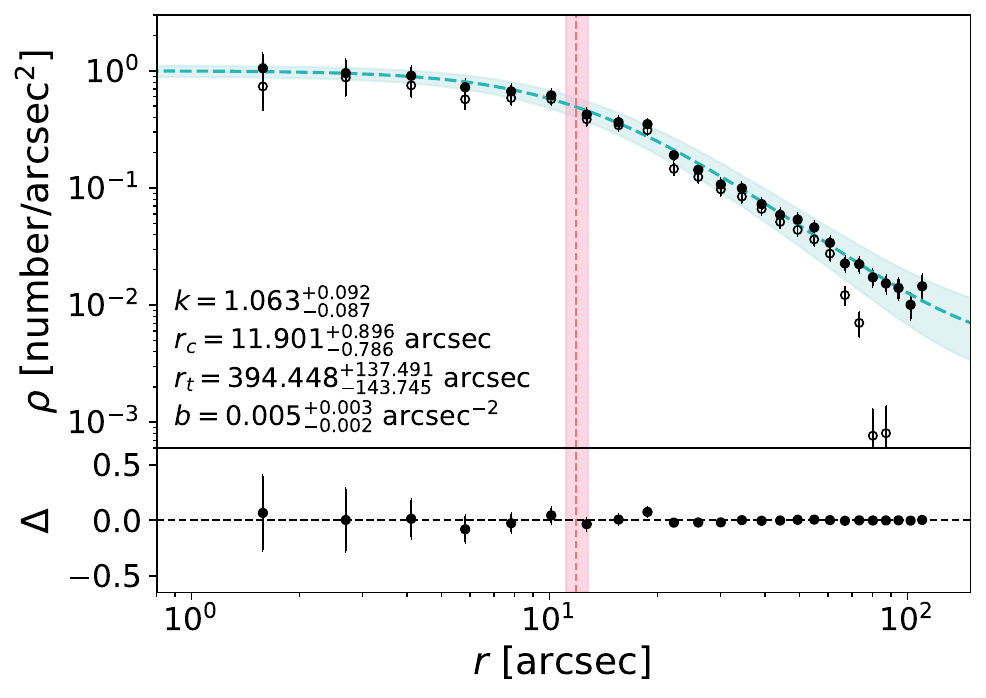}
 \caption{The radial number density profiles of NGC\,2173 are shown with filled circles representing the observed data and empty circles for the background-subtracted data. The dashed green curve is the best-fitting King model of the observations of NGC\,2173, and the green stripe marks the envelope of the $\pm 1\sigma$ solutions. The pink dashed vertical line marks the core radius of the best-fitting model, with the corresponding $\pm 1\sigma$ uncertainty represented by the pink stripe. The bottom panel shows the residuals between the best-fitting King model and the cluster density profile.}
 \label{f2}
 \end{figure}

Figure \ref{f1} outlines the procedure to select cluster members with F336W $\leq$ 23 mag, based on their PMs constrained to $\leq$ 0.14 mas/yr \citep[$\sim$ 33 km/s at the distance of NGC\,2173, distance modulus = 18.37 from][]{Milone2023}. This PM threshold is determined from the expected PM velocity dispersion of NGC\,2173, inferred from its total mass \citep[$5-7 \times 10^4 M_{\odot}$ from][]{Mclaughlin2005} and structural parameters (Figure \ref{f2}), calculated using Equation (1) in \citet{Fleck2006}. The PM selection effectively cleans the CMD, isolating spatially concentrated cluster members from background field stars, as demonstrated in Figure \ref{f1}. The selected stars with PMs similar to the cluster may also include field stars. To estimate the fraction of field stars contaminating the cluster member sample, we modeled the PM distribution of the raw stellar catalog using a best-fitting bi-Gaussian function. We found that most ($\sim$ 96\%) of stars with cluster-like PMs are indeed cluster members. This suggests that most stars in the BS region exhibited PMs consistent with cluster membership, are associated with NGC\,2173. Only one star in the BS region of the CMD was excluded due to PM inconsistency, indicating it is likely a field star.

We also fitted the empirical King model \citep{King1962}, with a constant background field population density, to the radial number density profiles of the raw stellar sample within the field of view of NGC\,2173:
\begin{equation}
\rho(r)=k\left[\frac{1}{\sqrt{1+\left(r / r_{\mathrm{c}}\right)^2}}-\frac{1}{\sqrt{1+\left(r_{\mathrm{t}} / r_{\mathrm{c}}\right)^2}}\right]^2+b.
\label{eq1}
\end{equation}
Where $r_{\rm c}$ and $r_{\rm t}$ are the core and tidal radii, respectively, $b$ is the background field number density, $k$ is a normalization coefficient, $\rho$ is the number density, and $r$ is the distance from a star to the cluster's center. Bright stars with F336W $\leq$ 22 mag, where photometric completeness exceeds 90\%, were taken into consideration. The best-fitting King model for the raw stellar sample in Figure \ref{f2} yielded a negligible background parameter $b$, indicating minimal contamination from field stars in the cluster's bright star population and supporting the robustness of the results presented by \citetalias[][]{Li2018} in the absence of field star subtraction. The cluster exhibits a flat core region with a core radius of 11.90 arcsec (2.86 pc). Our result is consistent with that derived by \citetalias[][]{Li2018}.

In Figure \ref{f2}, we also noticed that beyond 60 arcsec, the completeness of member stars declines significantly due to the reduced accuracy of PM measurements at the edges of the {\sl HST} images. The reason why stars with large radial distances have no PM determination in Figure \ref{f4}(a) depends on the fact that HST PMs are PMs relative to cluster members and when the distance from the cluster center is large we do not have enough bright cluster members to take as reference stars. This limitation predominantly affects stars in the outer regions of the cluster.

\subsection{Differential Extinction}

To mitigate the potential impact of spatially variable extinction, known as differential reddening, caused by variations in foreground interstellar gas and dust across the field of view, we applied a differential reddening correction using the method outlined in \citet{Milone2012}.
Briefly, a fiducial line of the main sequence (MS) was determined using a two-dimensional probability density function in the F336W versus F336W$-$F814W CMD. Bright MS stars were selected as the reference population to quantify the differential reddening affecting each star in the field of view. To streamline the correction process, the CMD was rotated counterclockwise to create a new reference frame, where the abscissa (${\rm X'}$) aligns with the reddening vector and the ordinate (${\rm Y'}$) is perpendicular to it. For each star, the average systematic offset ($\Delta {\rm X'}$) from the fiducial line was calculated based on the 50 nearest reference stars in space, excluding the target star. This offset served as the local differential reddening correction. The photometry was adjusted by subtracting $\Delta {\rm X'}$ from each star's ${\rm X'}$ value.

The corrected CMD was then used to refine the selection of reference stars and update the fiducial line. This iterative process, typically requiring about three iterations, was repeated until convergence. Finally, the corrected ${\rm X'}$ and ${\rm Y'}$ coordinates were transformed back into the original F336W and F814W magnitudes using the relative absorption coefficients from the extinction laws of \citet{Cardelli1989} and \citet{Odon1994}. This approach effectively minimizes the influence of differential reddening, enhancing the reliability of the subsequent analysis. While the reddening variation along the line of sight to this cluster is minor ($\Delta E(B-V) \sim$ 0.04 mag) and does not introduce visually noticeable changes to the CMD, the reddening-corrected CMD was still employed for all subsequent analyses.

\section{Results}\label{sec: results}

 \begin{figure}[ht!]
 \centering
\epsscale{1.15}
 \plotone{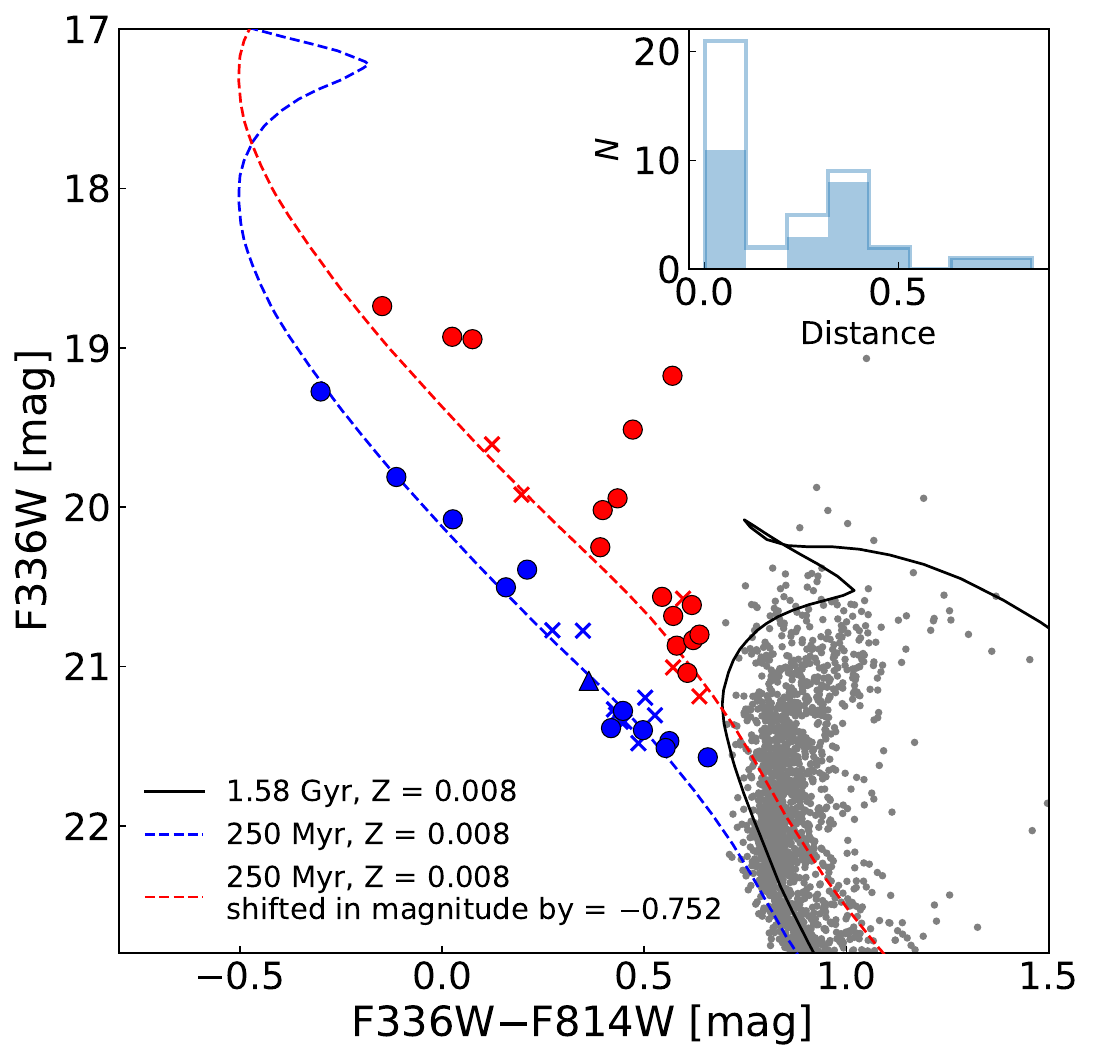}
 \caption{PM-selected CMD of NGC\,2173 zoomed into the BS region, with the BSs aligned along the blue- and the red-sequence plotted as blue and red dots, respectively. The black solid and blue dashed lines are best-fitting PARSEC isochrones with different ages labeled \citep{Bressan2012}. The red dashed line is the locus of the equal-mass binary sequence corresponding to the young isochrone. The crosses are candidates without reliable PM determination in the BS region of the CMD. The gray dots are PM-selected members. In the inset, the step and filled histograms represent the distributions of the perpendicular distances of all BS candidates and BS members from the single-star isochrone at the age of 250 Myr, respectively. Two well-defined peaks are visible.}
 \label{f3}
\end{figure}

\begin{figure}[ht!]
\centering
\epsscale{1.15}
\plotone{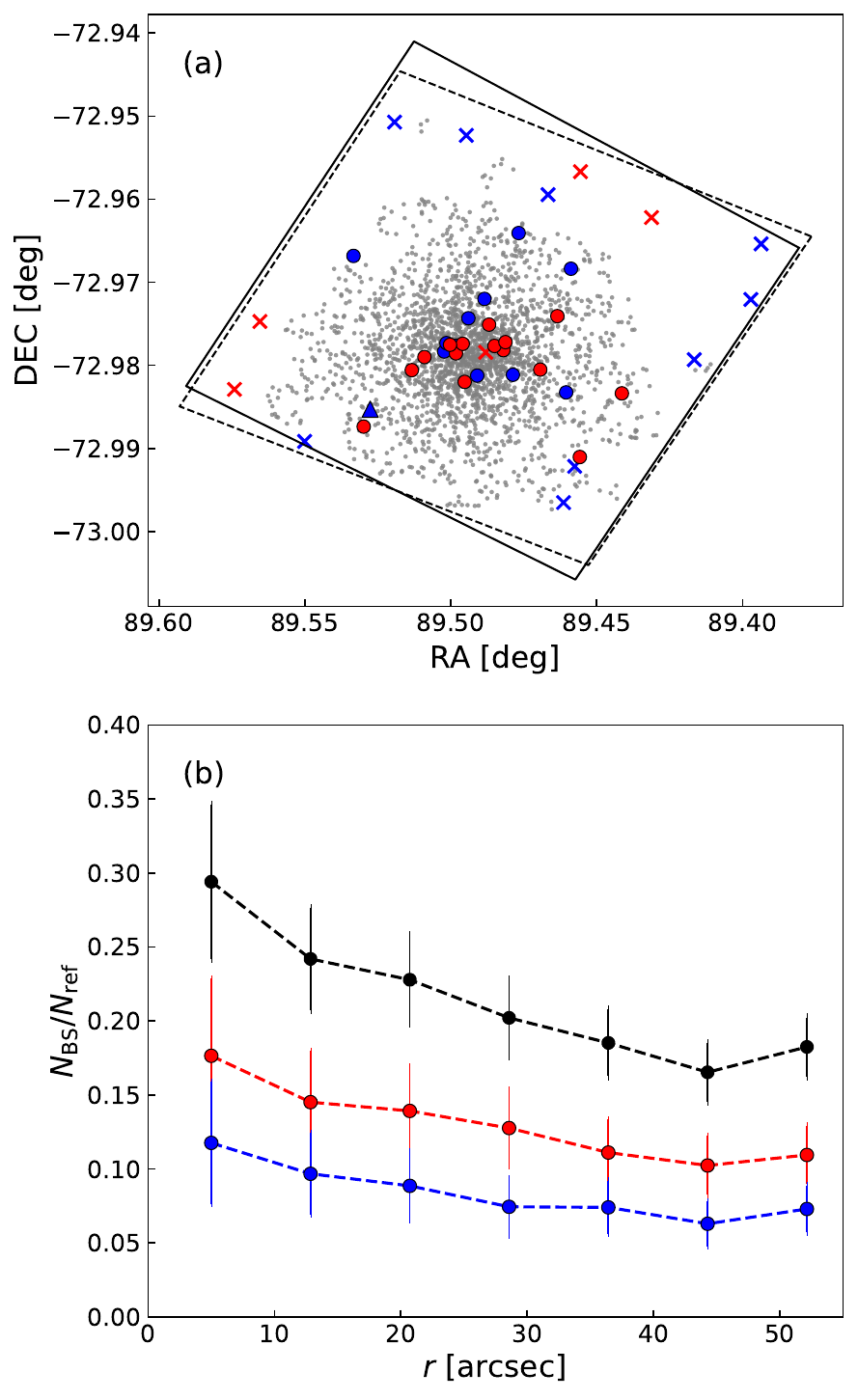}
\caption{(a) Spatial distributions of BSs in NGC\,2173. The black solid and dashed lines represent the fields for two-epoch observation in the F814W band. (b) BS-fraction profile normalized to the number of reference RGB and RC stars (black) as functions of the projected distance, $r$. The distributions of blue- (blue) and red-sequence BSs (red) are also plotted.}
\label{f4}
\end{figure}

The zoomed PM-selected CMD of NGC\,2173 in Figure \ref{f3} reveals an apparent bifurcated distribution of BSs, forming two distinct sequences consistent with similar patterns observed in previous old GCs. The classification of BSs into the red or blue sequence is determined by the perpendicular distance of each BS from the single-star isochrone at the age of 250 Myr. The BS candidates in the raw stellar catalog exhibit a more pronounced bifurcated pattern. However, the absence of reliable PM measurements for BSs near the image edges leaves their membership uncertain. The bifurcated sequence among the confirmed BS members around the cluster center is unquestionably evident. The numbers of the blue and red sequence BSs are comparable within 60 arcsec, with 11 BSs belonging to the blue sequence and 15 BSs associated with the red sequence. Due to the small number of members in NGC\,2173, as mentioned in Section \ref{sec: observations}, we should be cautious when using the PMs of distant stars in the outer regions of the cluster. When removing all stars with radial distances larger than 50 arcsec from the analysis, there are still 8 blue-sequence BSs and 13 red-sequence BSs.

We examined their radial distributions relative to the reference population---red giant branch (RGB) and red clump (RC) members---within a radius of 60 arcsec, where the reference sample has relatively high stellar completeness (Figure \ref{f2}). Figure \ref{f4} presents the stellar number ratios of all BSs, blue-sequence BSs, and red-sequence BSs to the reference stars. In NGC\,2173, the PM-selected BSs are slightly more centrally concentrated than the normal RGB and RC members, indicating that the cluster is at the dynamically young stage. The radial distributions of the blue and red sequences exhibit remarkably similar trends, showing consistent relative gradients across the cluster radius. We performed a statistical Kolmogorov-Smirnov test on the gradient distributions of the red and blue sequences, finding no significant difference between them ($p$-value = 0.96). 

Considering the expanded core region, it is unlikely that NGC\,2173 has undergone the core collapse event, unlike old GCs with similar bifurcated BS distributions. Collisional mechanisms are inefficient in producing such a significant coeval population of blue-sequence BSs in dynamically young clusters. Despite their distinct positions in the CMD, the similar radial behavior of the red and blue sequences suggests a common origin for both BS populations, likely driven by binary evolution. These findings support the hypothesis that the bifurcation in the CMD reflects variations within a single dominant formation pathway, i.e., binary evolution, rather than fundamentally different origins for the two sequences.

\section{Binary Population Synthesis} \label{sec: syntheticpop}

In the context of NGC\,2173, this section explores the critical role of binary MT systems in shaping the BS population in the unique environment of young clusters\footnote{The young clusters in this article refer to clusters younger than GCs, but not truly young clusters with ages $\textless$ 100 Myr.}, following the similar methodologies outlined in \citet{Jiang2017} and \citet{Jiang2022}.

\subsection{Binary Evolution Calculations}\label{sec: binarymodel}

\begin{figure*}[ht!]
\centering
\epsscale{1.15}
\plotone{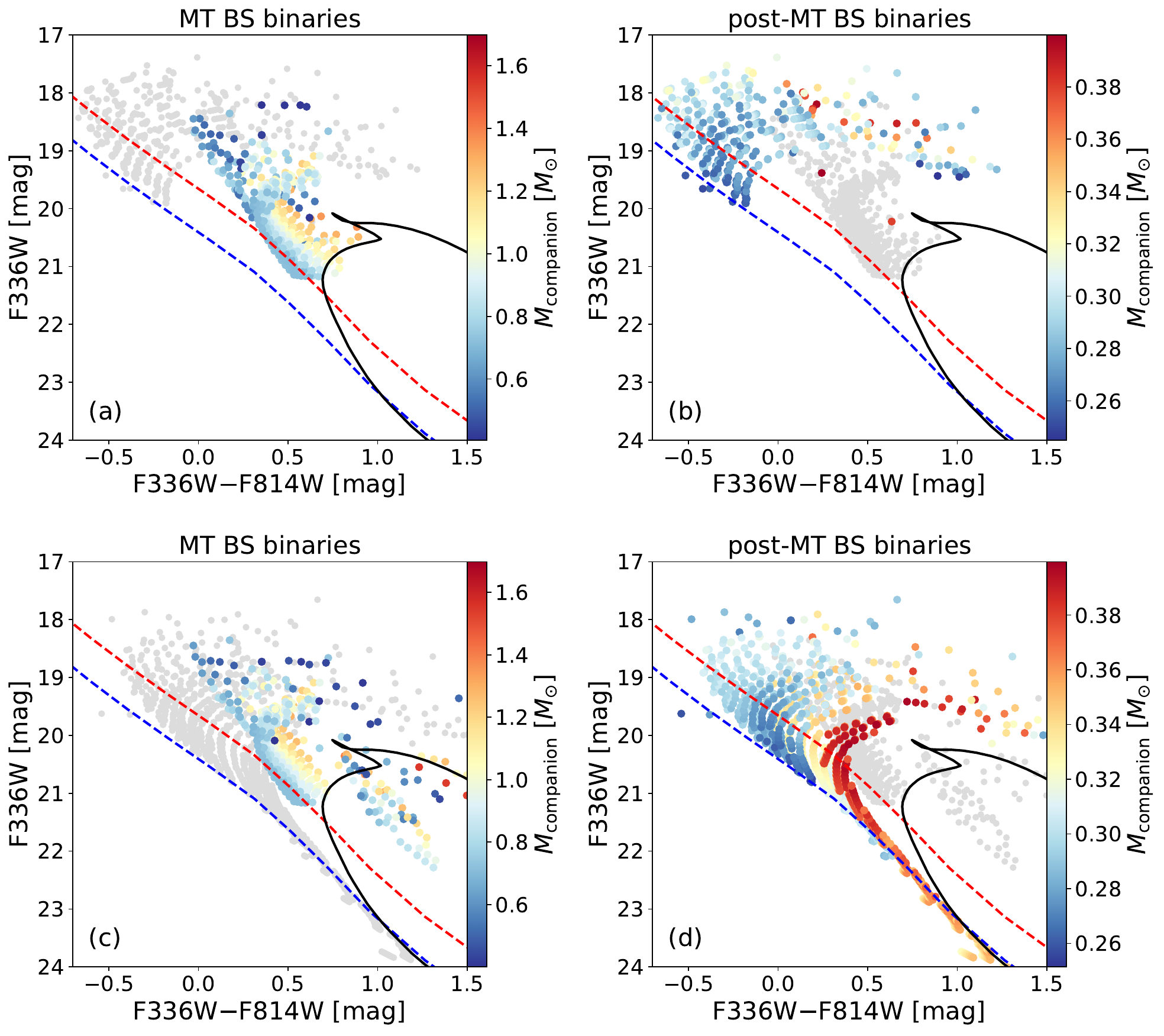}
\caption{(a) The locations of the unresolved MT and (b) post-MT BS binaries under conservative MT assumptions at 1.58 Gyr are shown in the F336W versus F336W$-$F814W CMD, with colors indicating the masses of their companions. For comparison, the positions of both types of unresolved BS binaries are overlaid as gray points in these panels. (c, d) Similar to panels (a, b), but under the segmented MT efficiency assumption.}
\label{f5}
\end{figure*}

To investigate the properties and distribution of BS binaries formed via binary evolution at the age of NGC\,2173, we used Eggleton's stellar evolution code (Program EV, private communication 2003)\footnote{This code can be obtained on request from ppe@igpp.ucllnl.org or Peter.Eggleton@yahoo.com.} in this study. This code, originally developed by \citet{Eggleton1971, Eggleton1972, Eggleton1973}, has been refined and updated over the past years \citep{Han1994, Pols1995, Nelson2001, Eggleton2002, Yakut2005, Eldridge2008}. We constructed a grid of binary evolutionary models with various combinations of initial primary mass $M_{10}$, the initial mass ratio $q_0$, and initial orbital period $P_0$ spacing at approximately equal logarithmic intervals:

\begin{equation}
\left\{
\begin{array}{l}
{\rm log}\,M_{10} = 0.230,\; 0.235,\; 0.240,\; \dots,\; 0.320,\\[1mm]
{\rm log}\,(1/q_0) = 0.050,\; 0.10,\; 0.15,\; \dots,\; 0.500,\\[1mm]
{\rm log}\,(P_0/P_{\rm ZAMS}) = 0.025,\; 0.050,\; 0.075,\; \dots,\; 1.000,
\end{array}
\right.
\label{eq2}
\end{equation}

where $P_{\rm ZAMS}$ refers to the orbital period at which the donor just fills its Roche lobe at the Zero Age Main Sequence (ZAMS) stage \citep{Nelson2001}. The parameter grid for binary evolution was meticulously constructed to encompass the parameter space for the progenitors of BS binaries at the age of NGC\,2173 as much as possible. As products of binary evolution, BS binaries in the two sequences arise from different initial binary parameters. Typically, the red sequence is primarily associated with Case A binary evolution, while the blue sequence predominantly originates from Case B binary evolution\footnote{Binary MT is classified into Cases A, B, and C based on the donor's evolutionary stage at its onset: MS, post-MS before core-helium depletion, and central helium burning and thereafter, respectively \citep{Kippenhahn1967}. Because stars expand during evolution, the initial orbital periods required for Case A, B, and C MT increase sequentially.} \citep{Lu2010, Jiang2017}. The MSTO mass of a 1.58 Gyr isochrone for NGC\,2173 is $\sim 1.7 M_{\odot}$. Therefore, the currently observable blue-sequence BSs formed through Case B MT in primordial binaries should originate from donor stars with initial masses roughly between 1.7 and 2.1 $M_{\odot}$. For the binaries with small initial mass ratios ($q_0  = M_{20}/M_{10}$), the MT rate can be extremely high and lead to dynamically unstable Roche lobe overflow (RLOF), leading to the formation of the common envelope. Since this study focuses on MT processes rather than coalescence, we only simulate binaries with $q_0 \geq q_{crit}$ \citep[adopting $q_{crit} = 0.32$ as in][]{chen2009}. We do not simulate binaries with ${\rm log}(P_0 / P_{\rm ZAMS}) \geq 1$, as in such cases, the donor star tends to expand significantly and fill its Roche lobe only in the late RGB phase, leading to Case C MT. The extremely high MT rate in Case C introduces various numerical challenges and complications in binary simulations, making it difficult to obtain reliable BS binary models formed through this channel. Therefore, this study primarily focuses on binaries undergoing stable Case A/B MT. The metallicity was set to $Z$ = 0.01, close to the metallicity of NGC\,2173 \citep{Milone2023}. When the donor fills its Roche lobe, its material is transferred to the accretor through RLOF. We simply assumed circular orbits for all binaries. Tides were neglected, allowing stellar rotation\footnote{The initial rotation rates of the stars are assigned based on their initial masses as in \citet{Hurley2000}.} and orbital rotation to evolve independently. This continues until RLOF occurs, at this point, synchronization is enforced, transferring angular momentum between the stellar spin and the orbit. Convective overshooting is taken into account as in \citet{pols1998} with a specified constant $\delta_{\rm ov} = 0.12$. Since stellar-wind mass loss during early evolutionary stages (e.g., the core hydrogen-burning phase) is relatively minor compared to RLOF or the stellar-wind mass loss in later evolutionary stages \citep{De2007}, it is not considered in this study.

The evolutionary calculations of these binaries were terminated when both components filled their Roche lobes or the code failed to converge. The binaries evolved from the ZAMS to the cluster's age of 1.58 Gyr. We converted the theoretical parameters of both components of these BS binaries into the {\sl HST}'s UVIS/WFC3 filters absolute magnitudes by using the PARSEC Bolometric Correction\footnote{\url{https://sec.center/YBC}} ({\tt\string YBC}, \citealt{Chen2019}), facilitating a direct comparison with the observations of NGC\,2173. In our study, the binaries are assumed to be unresolved, and their combined magnitudes are calculated using the following formula:
\begin{equation}
m_{\rm binary} = m_{\rm BS} - 2.5 {\rm log}(1 + 10^{(m_{\rm BS} -m_{\rm companion})/2.5}),
\label{eq3}
\end{equation}
where $m_{\rm BS}$ and $m_{\text{companion}}$ are the magnitudes of the BS components (accretors) and their companions (donors) in unresolved BS binaries, respectively. 

Under the idealized conservative assumption, where the mass and angular momentum of the binaries are assumed to be conservative during the MT phase, the results are presented in Figure \ref{f5}(a, b). Ongoing MT produces BS binaries exclusively in the red-sequence region. In these MT BS binaries, the donor (initially the more massive star) is still transferring mass to the accretor (initially the less massive star), but the current mass ratio has already reversed. The accretor becomes a BS, while the donor, despite similar luminosity, appears much redder. As BSs and their companions cannot be resolved, these unresolved binaries appear brighter and redder than single BSs, placing them in the red sequence region. In contrast, post-MT systems populate both the red- and blue-sequence areas, depending on the evolutionary stages of the donor and accretor. After the MT ends, the position of the BS-WD binary in the optical CMD depends on the WD's cooling timescale and the remaining MS lifetime of the BS component. If the cooling timescale is much shorter than the BS's remaining lifetime on the MS, the binary will enter the blue sequence region, as the WD contributes negligibly to the binary's magnitude. However, if the WD has not sufficiently cooled or the BS is close to evolving off the MS, the system will remain in or transition to the red sequence region. These distributions are similar to the results obtained in old GCs (10--15 Gyr) as reported by \citet{Jiang2017} and \citet{Jiang2022}.

However, conservative MT calculations fail to produce low-mass BS binaries in the lower part of the blue sequence as shown in Figure \ref{f5}. This limitation is particularly evident in younger star clusters but becomes negligible in older GCs, as seen in comparison with \citet{Jiang2022}. The primary reason is that in younger clusters, the MSTO shifts downward too rapidly in the CMD compared to older clusters. In our initial tests of conservative binary evolution, we also calculated the evolution of lower-mass primordial binaries (i.e., those with initial donor masses below the MSTO mass $\sim 1.7 M_{\odot}$). We found that they still failed to populate the lower part of the blue sequence, see Appendix \ref{sec: append} for details. To address the lack of low-mass blue-sequence BSs, the more realistic and complex non-conservative evolution, in which a fraction of the masses and orbital angular momentums are lost from the systems, are taken into account. One of the major uncertainties in close binary evolution is the efficiency of MT ($\beta$), defined as the mass fraction of the matter lost from the donor accreted by the accretor. \citet{De2007} suggested that MT efficiency decreases with increasing initial orbital period, based on observations of massive double-lined eclipsing binaries. This correlation can be explained by the critical role of tidal interactions in influencing accretor spin within binary systems. The accretor efficiently accumulates the transferred material along with the associated angular momentum, resulting in spin-up. Once the accretor reaches critical spin, accretion halts, marking a transition from conservative to non-conservative MT. In short-period binaries, tidal interactions can mitigate the impact of spin-up, allowing the accretor to absorb more mass before reaching critical rotation. In contrast, long-period Case B MT binaries typically exhibit highly non-conservative MT processes \citep{Wang2022}. However, \citet{Popham1991} suggested that accretors may continue to gain mass even after reaching breakup spin by transferring angular momentum outward vi accretion disks.

\begin{figure*}[ht!]
\centering
\epsscale{1.15}
\plotone{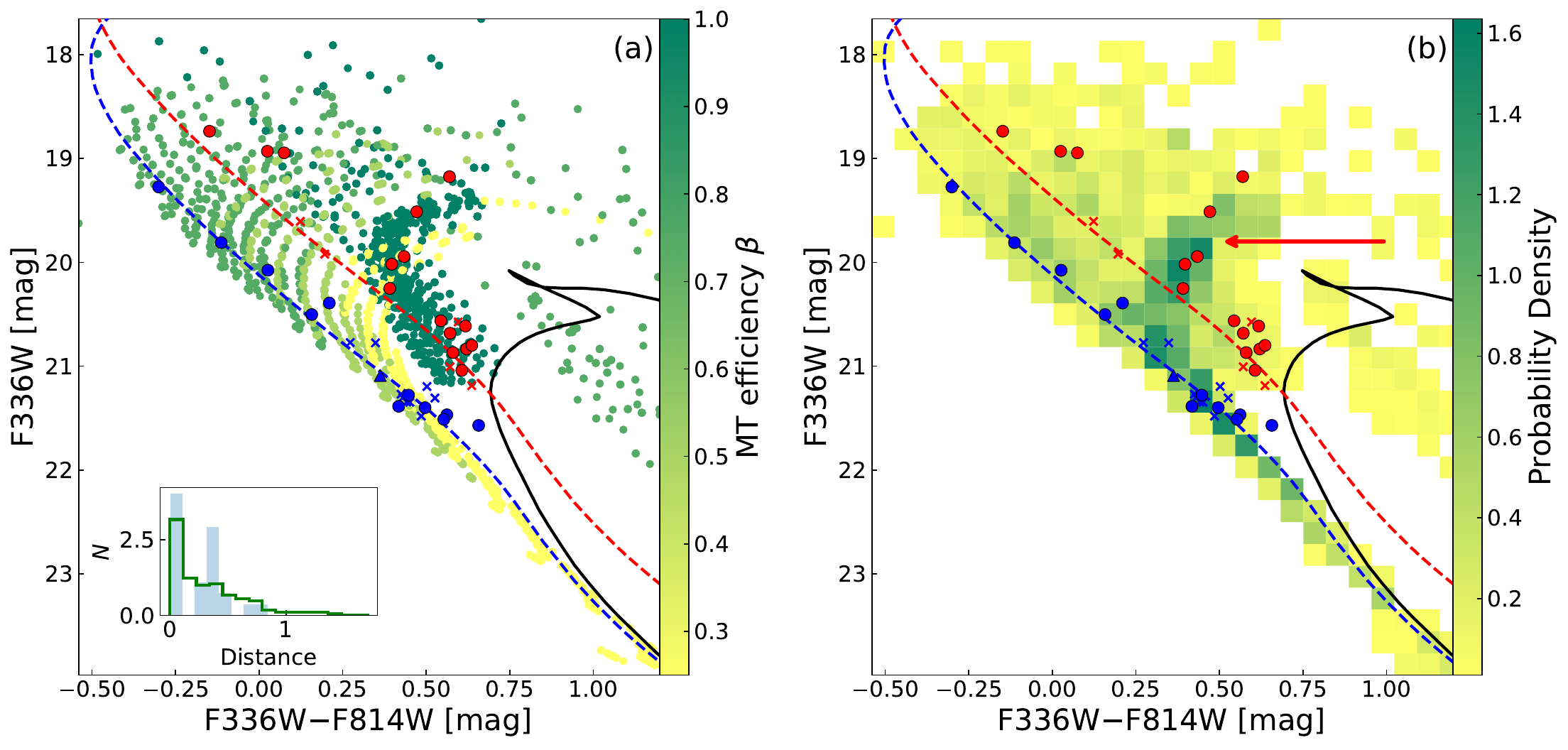}
\caption{(a) Monte Carlo simulation of binary-origin BSs compared to the observed BSs in NGC\,2173, with colors representing the MT efficiency ($\beta$). The inset shows normalized distributions: the step histogram represents the perpendicular distances of simulated BS binaries from the single-star isochrone at the age of 250 Myr, while the filled histogram corresponds to BS members in NGC\,2173. (b) Similar to panel (a), but with colors indicating the probability density map for BS formation, overlaid with the positions of simulated BS binaries. The red arrow indicates the positions of MT BS binaries under the assumption of conservative MT, as shown in Figure \ref{f5}(a).}
\label{f6}
\end{figure*}

We adopted a simplified segmented MT efficiency assumption where $\beta$ decreases as the initial orbital period $P_0$ increases, according to \citet[their Figure 7]{De2007},
\begin{equation}
\beta = \left\{
\begin{array}{ll}
100\%, & \mbox{if } \log P_0 < 0.1, \\[1mm]
75\%,  & \mbox{if } 0.1 \le \log P_0 < 0.3, \\[1mm]
50\%,  & \mbox{if } 0.3 \le \log P_0 < 0.5, \\[1mm]
25\%,  & \mbox{if } 0.5 \le \log P_0.
\end{array}
\right.
\label{eq4}
\end{equation}
Here the unit of $P_0$ is day. We note that the above relation should be treated with caution. The dependence of $\beta$ on $P_0$ is just based on the study of more massive binaries \citep[7-25 $M_{\odot}$;][]{De2007}, where the derived $\beta$ values show significant uncertainty. Our assumption of $\beta$ in this work is merely one possible way it may vary in the parameter space. $\beta = 100\%$ represents conservative MT, where the companion star fully accretes all material lost from the donor. In contrast, $\beta < 100\%$ indicates non-conservative MT, where the mass lost from the binary system carries away specific angular momentum. In the code of Eggleton, the escape of the mass can occur in the form of bipolar jets moving from the system. Compared to BS binaries in the blue region with the same initial donor masses, those in the red region tend to have shorter initial orbital periods. Therefore, modifications to the MT efficiency primarily affect BS binaries in the blue sequence. 

Figure \ref{f5}(c, d) presents the results under the segmented MT efficiency assumption, successfully reproducing low-mass BS systems in the blue region by incorporating mass loss during MT. This suggests that non-conservative MT is necessary for forming low-mass blue BSs in young clusters through binary interactions. In Figure \ref{f5}(b, d), we find that the relationship between the positions of post-MT BS binaries on the CMD and their WD companion masses differs significantly from the results obtained under the conservative MT assumption. In the conservative MT cases, the faintest post-MT BS binaries originate from our grid's least massive primordial binaries $1.7(1+q_{crit})M_{\odot}$ and host the least massive helium WDs. However, under the segmented MT efficiency assumption, the faint post-MT BS binaries host more massive WDs, evolving from the most massive initial donors (2.1 $M_{\odot}$), contrary to the conservative case. This discrepancy is strongly dependent on the MT efficiency assumption we adopted: the larger the donor's initial mass, the larger its initial orbital period for Case B MT, typically resulting in a lower MT efficiency. Material accreted by the accretor decreases, which causes a decrease in luminosity when compared with the conservative case.

\subsection{Comparison with NGC\,2173}

To compare theoretical predictions with observations, we performed a Monte Carlo simulation of a star cluster based on a grid of binary evolution models under the segmented MT efficiency assumption to conduct a comprehensive BS binary population synthesis, facilitating the exploration of various properties, such as the number ratio of BSs in different regions in the CMD. In this study, the synthetic BS population was modeled by incorporating empirical distributions of initial parameters: (1) the initial mass function (IMF) from \citet{Miller1979}; (2) a constant mass-ratio distribution ($n(q_0) = 1$); (3) a flat logarithmic distribution of the initial orbital separation ($a_0$).

As illustrated in Figure \ref{f6}, most observed BSs in NGC\,2173 fall within the theoretical regions predicted for the two types of BS binaries. Most of the observed BSs tend to cluster in areas of higher probability density as predicted by the simulation, specifically (1) along the ZAMS, corresponding to the presumed blue sequence, where the number of BS binaries is highest (see inset in Figure \ref{f6}(a)), and (2) in the region indicated by the red arrow in Figure \ref{f6}(b). Region (2), located in the middle or lower part of the red-sequence BS region, is where MT BS binaries under the conservative MT assumption are concentrated as shown in Figure \ref{f5}(a). It is worth noting that the theoretical predictions only produce a gap at the lower part of the BS region, which does not entirely align with the observed distinct gap between the blue and red sequences. The predicted fractions of BSs in different regions of the CMD can serve as a critical test for binary evolution scenarios when compared with observed fractions. Taking Poisson uncertainties into account, the observed ratio of BSs in the different sequences of NGC\,2173 ($N_{\rm{blue BS}} / N_{\rm{red BS}} = 8/13 \approx 0.62$) is in agreement with the predicted value of approximately 0.65 from binary population synthesis.

Some observed low-mass BSs in the blue sequence are located within the theoretical gap between the two types of BS binaries. As discussed in Section \ref{sec: binarymodel}, this could be the result of limitations in our binary model grid, such as the uncertainties in MT efficiency. Leaving these uncertainties aside, these BSs may have distinct origins. As discussed in \citet{Xin2015} and \citet{Jiang2022}, single BSs formed through mergers offer a theoretical explanation for forming these BSs within the observed gap. This is because BSs formed through binary mergers inherit cores resembling those of their more evolved progenitors, whereas BSs formed through MT do not. A rough estimate based on the CMD suggests that at least two observed BSs in the gap likely require binary mergers to explain their formation.

Figure \ref{f6}(a) also illustrates the relationship between MT efficiency and the CMD positions of BS binaries. The blue sequence primarily consists of systems formed under non-conservative MT conditions (mostly with $\beta < 1$), while the red sequence predominantly represents systems formed under conservative MT conditions. Our results highlight the critical role of MT efficiency in shaping the BS population in CMDs. In particular, the blue sequence's reliance on non-conservative MT underscores the importance of realistic assumptions about mass loss in binary evolution models. 

\section{Discussion} \label{sec: discussion}

In this study, the main findings of our background subtraction, combined with our previous results on the LMC cluster NGC\,1783 \citep{Wang2024}, emphasize the need for caution when classifying stars as field stars based solely on their spatial distribution in {\sl HST} observations. The limited field of view of {\sl HST} can sometimes give a misleading impression resulting in the misclassification of cluster-associated stars as field stars and leading to erroneous analyses \citep{Cabrera2016, Dalessandro2019a, Dalessandro2019b}. For MC clusters, kinematic information remains virtually the only reliable method for field star subtraction.

We identified an apparent bifurcated distribution of BSs in NGC\,2173, resembling the double sequence observed in the core-collapsed GC M\,30. However, the density profile of NGC\,2173 does not exhibit the central cusp typically associated with core collapse. To further investigate the origin of these BSs, we performed a realistic simulation using the high-performance $N$-body code {\tt\string PETAR}\footnote{\url{https://github.com/lwang-astro/PeTar}} \citep{Wang2020, WangL2022}, similar to the configuration in \citet{Wang2024}, with the simulated cluster having comparable age, total mass, and structural parameters to NGC\,2173. No BSs formed through direct stellar collisions were found in the simulated cluster. Based on observational evidence and numerical results, the expected number of collision-induced BSs in non-core-collapsed clusters is negligible, ruling out stellar collisions as the origin of the observed tight blue sequence BSs. \textit{Binary evolution is the prominent pathway to form BSs}. Collisions may produce BSs, but mainly in a limited number of core-collapsed clusters, as well as dynamically unstable triple or higher-order systems \citep[e.g.,][]{perets2012}.

We calculated a grid of binary evolution models appropriate for the BSs currently observed in NGC\,2173. Our analysis revealed that conservative models cannot produce fainter blue-sequence BSs because, in the binary MT scenario, blue-sequence BS binaries require the cessation of Case B MT. The donor's mass must exceed the MSTO mass, and the accretor's initial mass must be sufficient to undergo stable RLOF. Under conservative conditions, the relatively massive accretor absorbs all the mass lost by the donor as it evolves into a helium WD. Consequently, BS binaries formed under conservative MT always have luminosities above the MSTO. Conservative MT represents an extreme case in binary evolution, producing the brightest BSs in the blue region, especially in such young clusters.

Clearly, if the companion star retains only a small fraction of the transferred mass, the resulting increase in mass and luminosity will be limited. Non-conservative MT is essential for shaping the distribution of post-MT BS binaries in the CMD, as shown in Figure \ref{f5}. \citet{Sun2021, Sun2023} explore the detailed models that best reproduce the primary characteristics of two BS-WD binaries (WOCS 5379 and WOCS 4540) in the OC NGC\,188 (6--7 Gyr) while allowing for the possibility of non-conservative MT. This non-conservative MT is key to expanding the orbit fast enough to permit stable MT. An interesting scientific implication appears in Figure \ref{f5}(d): low-mass post-MT BS binaries merging with the MS in the CMD. These systems, formed under the highly non-conservative assumption, all have WD companions, with 35\% of them exhibiting orbital periods of $\geq$100 day. Their locations align with the so-called ``blue lurkers (BLs)'', considered low-luminosity counterparts of BSs. BLs are a newly discovered class of stars in the OC M\,67, observationally identified as unusually fast rotators in wide binaries \citep{Leiner2019}. The inefficient MT is essential for explaining the observed BLs. If the accretors were to accrete excessive material, they would evolve beyond the MSTO and transform into BSs. Two BLs with WD companions in M\,67 have been identified, with binary evolution models yielding plausible evolutionary pathways to both BLs via highly non-conservative MT \citep{Nine2023, Sun2024}. This qualitative consistency between simulations and observations underscores that realistic MT efficiencies are essential for reproducing the observed low-mass BSs and BLs in young clusters. 

Binary evolution simulations reveal that the distribution of BSs is nearly continuous, with gaps only appearing at the low-mass ends of the blue and red sequences. Incorporating effects such as unstable MT or binary mergers results in a more complex and nearly continuous distribution in the CMD \citep{Jiang2017}. Considering the observed ratio of BSs in the different sequences of NGC\,2173 consistent with the predicted value from binary population synthesis, the limited number of BSs in NGC\,2173 raises the question of whether the observed two-sequence phenomenon in young clusters could result from statistical biases due to small sample sizes. By randomly selecting the same number of BSs as observed in NGC\,2173 from the Monte Carlo simulations and repeating this process 10,000 times, the statistical probability of reproducing a bifurcated BS distribution closely matching the observed feature was only 2\%. Using the catalog from \citet{Milone2023}, we find that among star clusters in the MCs with ages comparable to NGC\,2173, the occurrence rate of bifurcated sequences of BSs is 1/20 = 5\%. This observed frequency is of the same order as the 2\% prediction from our simulations. In young clusters like NGC\,2173, the observed bifurcation of BSs is more likely a statistical coincidence rather than reflecting the intrinsic difference in formation mechanisms as proposed by \citet{Ferraro2009}.

For old GCs, theoretical analyses within the binary evolution framework suggest that the distribution of MT and post-MT BSs resembles predictions for young clusters, with blue sequence BSs primarily originating from primordial binaries that have completed MT \citep{Jiang2017, Jiang2022}. The number of these evolutionary BSs formed mostly from compact binary evolution depends on initial binary properties. After the core collapse, dynamical interactions become significant. According to \citet{Hypki2013}, which examined BS formation through different channels in star clusters, only a few BSs were found to result from the direct collision of two single stars. The number of direct collisional BSs does not significantly increase even after the core collapse, with most collisional BSs instead forming in three- and four-body interactions in the cluster core. These interactions often impart high escape velocities, leading to most dynamically formed BSs escaping the cluster and appearing as high-velocity stars in the galactic field \citep[e.g.,][]{Tillich2010}. We therefore speculate that the contribution of collision-induced BSs to the blue sequence is not substantial as suggested by \citet{Ferraro2009}. This topic lies beyond the scope of the current study and will be addressed in future work.

Our binary evolution models predict that 70\% of the BS binaries are BS+WD binary systems. According to \citet{Han1995}, these systems are expected to undergo the second MT phase, with a 20\% probability of naturally evolving into WD+WD binaries. These systems, when inspiraling, are important objects for space-borne gravitational-wave detectors in the millihertz frequency bands \citep{Ren2023}.

\section{Summary} \label{sec: summary}

Based on multi-epoch {\sl HST} data and binary population synthesis, this study examines the bifurcated BS sequences in NGC\,2173. Our main results and conclusions are summarized below:
\begin{enumerate}[label=(\arabic*)]
 \item The overall distribution of PM-selected BSs in NGC\,2173 draws a well-defined narrow blue sequence and a more sparse red sequence, consistent with \citetalias[][]{Li2018}.
 \item BS binaries formed through the MT can populate the CMD region of observed BSs at the age of NGC\,2173. The blue sequence comprises BS+WD binaries that had finished non-conservative MT, while the red sequence predominantly consists of ongoing MT binaries under conservative assumption.
 \item Star clusters may contain a substantial number of WD+WD binaries.
\end{enumerate}

\begin{acknowledgements}
We thank the anonymous referee for the valuable comments and suggestions for improving our manuscript.
We acknowledge Zhongjie Zheng at the Guangdong Ocean University, People's Republic of China, for his insightful suggestions. 
L. W. and C. L. are supported by the National Natural Science Foundation of China (NSFC) through grant 12233013. 
D.J. acknowledges support from the NSFC (Nos. 12288102, 12473033, 12073070, 12333008, 12090040/3), the National Key R\&D Program of China No. 2021YFA1600403, the Western Light Project of CAS (No. XBZG-ZDSYS-202117), International Centre of Supernovae, Yunnan Key Laboratory (No. 202302AN360001), Yunnan Revitalization Talent Support Program -- Science \& Technology Champion Project (NO. 202305AB350003), Yunnan Fundamental Research Project (No. 202201BC070003, 202401BC070007), and the Yunnan Ten Thousand Talents Plan Young \& Elite Talents Project. 
This work has received funding from ``PRIN 2022 2022MMEB9W - {\it Understanding the formation of globular clusters with their multiple stellar generations}" (PI Anna F.\,Marino), from INAF Research GTO-Grant Normal RSN2-1.05.12.05.10 - (ref. Anna F.\,Marino) of the "Bando INAF per il Finanziamento della Ricerca Fondamentale 2022", and from the European Union's Horizon 2020 research. 
L.W. is grateful for support from both the NSFC through grant 21BAA00619 and the one-hundred-talent project of Sun Yat-sen University.
\end{acknowledgements}

\appendix

\section{Case A MT in lower-mass binaries \label{sec: append}}

\begin{figure*}[ht!]
\centering
\epsscale{1.15}
\restartappendixnumbering
\plotone{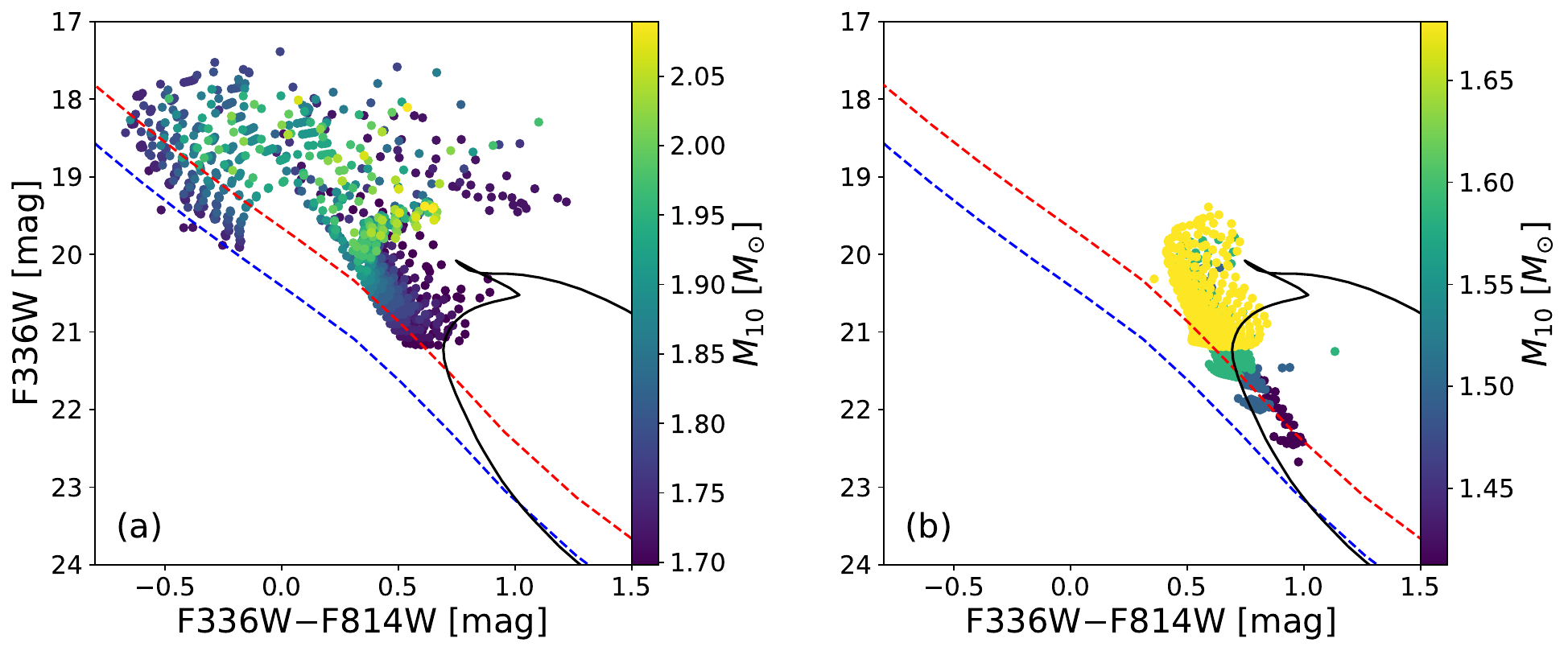}
\caption{(a) The distribution of the unresolved BS binaries at 1.58 Gyr in the CMD under the assumption of conservative MT, with initial donor masses above 1.7 $M_{\odot}$. The colors indicate the initial masses of the donors ($M_{10}$). (b) Similar to panel (a), but for the unresolved BS binaries with initial donor masses below 1.7 $M_{\odot}$ (${\rm log}M_{10} = 0.150, 0.175, 0.200, 0.225$).}
\label{fA1}
\end{figure*}

In this study, we did not consider the possibility of stable Case A MT in lower-mass primordial binaries (i.e., those with initial donor masses below the MSTO mass $\sim 1.7 M_{\odot}$; ${\rm log}M_{10} = 0.150, 0.175, 0.200, 0.225$). Under the assumption of conservative MT, we found that these binaries produce only red-sequence BSs at the low-mass end and do not contribute to the blue sequence, as shown in Figure \ref{fA1}. At the current cluster age (1.58 Gyr), these primordial binary systems can only undergo Case A MT, which is often unstable and can lead to contact or merger--especially for systems with higher initial donor masses, smaller mass ratios, and shorter orbital periods. Thus, a significant fraction of close binary models did not survive. Since the donor undergoes minimal expansion during its MS evolution, Case A MT only starts relatively late and closer to the current cluster age. This results in an insufficient time interval between the onset of MT and the current cluster age, meaning that all these systems are still undergoing MT at the cluster's current age. Some accretors have not yet gained enough mass to enter the BS region and may evolve into primordial BSs later. Some have already reached the BS region but remain in the red sequence since MT is still ongoing. Consequently, primordial binaries with initial donor masses below the MSTO cannot populate the lower part of the blue sequence at this time.

\vspace{5mm}


\bibliography{NGC2173}{}

\begin{thebibliography}{}
\expandafter\ifx\csname natexlab\endcsname\relax\def\natexlab#1{#1}\fi
\providecommand{\url}[1]{\href{#1}{#1}}
\providecommand{\dodoi}[1]{doi:~\href{http://doi.org/#1}{\nolinkurl{#1}}}
\providecommand{\doeprint}[1]{\href{http://ascl.net/#1}{\nolinkurl{http://ascl.net/#1}}}
\providecommand{\doarXiv}[1]{\href{https://arxiv.org/abs/#1}{\nolinkurl{https://arxiv.org/abs/#1}}}

\bibitem[{Andronov {et~al.}(2006)Andronov, Pinsonneault, \&
  Terndrup}]{Andronov2006}
Andronov, N., Pinsonneault, M.~H., \& Terndrup, D.~M. 2006, \apj, 646, 1160,
  \dodoi{10.1086/505127}

\bibitem[{Beccari {et~al.}(2019)Beccari, Ferraro, Dalessandro, Lanzoni, Raso,
  Origlia, Vesperini, Hong, Sills, Dieball, \& Knigge}]{Beccari2019}
Beccari, G., Ferraro, F.~R., Dalessandro, E., {et~al.} 2019, \apj, 876, 87,
  \dodoi{10.3847/1538-4357/ab13a4}

\bibitem[{Billi {et~al.}(2023)Billi, Ferraro, Mucciarelli, Lanzoni, Cadelano,
  Monaco, Mateo, Bailey, Reiter, \& Olszewski}]{Billi2023}
Billi, A., Ferraro, F.~R., Mucciarelli, A., {et~al.} 2023, \apj, 956, 124,
  \dodoi{10.3847/1538-4357/acf372}

\bibitem[{{Bodensteiner} {et~al.}(2021){Bodensteiner}, {Sana}, {Wang},
  {Langer}, {Mahy}, {Banyard}, {de Koter}, {de Mink}, {Evans}, {G{\"o}tberg},
  {Patrick}, {Schneider}, \& {Tramper}}]{Bodensteiner2021}
{Bodensteiner}, J., {Sana}, H., {Wang}, C., {et~al.} 2021, \aap, 652, A70,
  \dodoi{10.1051/0004-6361/202140507}

\bibitem[{{Bodensteiner} {et~al.}(2023){Bodensteiner}, {Sana}, {Dufton},
  {Wang}, {Langer}, {Banyard}, {Mahy}, {de Koter}, {de Mink}, {Evans},
  {G{\"o}tberg}, {H{\'e}nault-Brunet}, {Patrick}, \&
  {Schneider}}]{Bodensteiner2023}
{Bodensteiner}, J., {Sana}, H., {Dufton}, P.~L., {et~al.} 2023, \aap, 680, A32,
  \dodoi{10.1051/0004-6361/202345950}

\bibitem[{Bressan {et~al.}(2012)Bressan, Marigo, Girardi, Salasnich, Dal~Cero,
  Rubele, \& Nanni}]{Bressan2012}
Bressan, A., Marigo, P., Girardi, L., {et~al.} 2012, \mnras, 427, 127,
  \dodoi{10.1111/j.1365-2966.2012.21948.x}

\bibitem[{Cabrera-Ziri {et~al.}(2016)Cabrera-Ziri, Niederhofer, Bastian,
  Rejkuba, Balbinot, Kerzendorf, Larsen, Mackey, Dalessandro, Mucciarelli,
  Charbonnel, Hilker, Gieles, \& Hénault-Brunet}]{Cabrera2016}
Cabrera-Ziri, I., Niederhofer, F., Bastian, N., {et~al.} 2016, \mnras, 459,
  4218, \dodoi{10.1093/mnras/stw966}

\bibitem[{{Cardelli} {et~al.}(1989){Cardelli}, {Clayton}, \&
  {Mathis}}]{Cardelli1989}
{Cardelli}, J.~A., {Clayton}, G.~C., \& {Mathis}, J.~S. 1989, \apj, 345, 245,
  \dodoi{10.1086/167900}

\bibitem[{Chen \& Han(2009)}]{chen2009}
Chen, X., \& Han, Z. 2009, \mnras, 395, 1822,
  \dodoi{10.1111/j.1365-2966.2009.14669.x}

\bibitem[{Chen {et~al.}(2019)Chen, Girardi, Fu, Bressan, Aringer, Dal~Tio,
  Pastorelli, Marigo, Costa, \& Zhang}]{Chen2019}
Chen, Y., Girardi, L., Fu, X., {et~al.} 2019, \aap, 632, A105,
  \dodoi{10.1051/0004-6361/201936612}

\bibitem[{Clarkson {et~al.}(2011)Clarkson, Sahu, Anderson, Rich, Smith, Brown,
  Bond, Livio, Minniti, Renzini, \& Zoccali}]{Clarkson2011}
Clarkson, W.~I., Sahu, K.~C., Anderson, J., {et~al.} 2011, \apj, 735, 37,
  \dodoi{10.1088/0004-637X/735/1/37}

\bibitem[{Cordoni {et~al.}(2023)Cordoni, Milone, Marino, Vesperini, Dondoglio,
  Legnardi, Mohandasan, Carlos, Lagioia, Jang, \& Ziliotto}]{Cordoni2023}
Cordoni, G., Milone, A.~P., Marino, A.~F., {et~al.} 2023, \aap, 672, A29,
  \dodoi{10.1051/0004-6361/202245457}

\bibitem[{{Dalessandro} {et~al.}(2019{\natexlab{a}}){Dalessandro}, {Ferraro},
  {Bastian}, {Cadelano}, {Lanzoni}, \& {Raso}}]{Dalessandro2019a}
{Dalessandro}, E., {Ferraro}, F.~R., {Bastian}, N., {et~al.}
  2019{\natexlab{a}}, \aap, 621, A45, \dodoi{10.1051/0004-6361/201834011}

\bibitem[{{Dalessandro} {et~al.}(2019{\natexlab{b}}){Dalessandro}, {Ferraro},
  {Bastian}, {Cadelano}, {Lanzoni}, \& {Raso}}]{Dalessandro2019b}
---. 2019{\natexlab{b}}, RNAAS, 3, 38, \dodoi{10.3847/2515-5172/ab0829}

\bibitem[{Dalessandro {et~al.}(2013)Dalessandro, Ferraro, Massari, Lanzoni,
  Miocchi, Beccari, Bellini, Sills, Sigurdsson, Mucciarelli, \&
  Lovisi}]{Dalessandro2013}
Dalessandro, E., Ferraro, F.~R., Massari, D., {et~al.} 2013, \apj, 778, 135,
  \dodoi{10.1088/0004-637X/778/2/135}

\bibitem[{{de Mink} {et~al.}(2007){de Mink}, {Pols}, \& {Hilditch}}]{De2007}
{de Mink}, S.~E., {Pols}, O.~R., \& {Hilditch}, R.~W. 2007, \aap, 467, 1181,
  \dodoi{10.1051/0004-6361:20067007}

\bibitem[{Eggleton(1971)}]{Eggleton1971}
Eggleton, P.~P. 1971, \mnras, 151, 351, \dodoi{10.1093/mnras/151.3.351}

\bibitem[{Eggleton(1972)}]{Eggleton1972}
---. 1972, \mnras, 156, 361, \dodoi{10.1093/mnras/156.3.361}

\bibitem[{Eggleton(1973)}]{Eggleton1973}
---. 1973, \mnras, 163, 279, \dodoi{10.1093/mnras/163.3.279}

\bibitem[{Eggleton \& Kiseleva-Eggleton(2002)}]{Eggleton2002}
Eggleton, P.~P., \& Kiseleva-Eggleton, L. 2002, \apj, 575, 461,
  \dodoi{10.1086/341215}

\bibitem[{{Ekanayake} \& {Wilhelm}(2018)}]{Ekanayake2018}
{Ekanayake}, G., \& {Wilhelm}, R. 2018, \mnras, 479, 2623,
  \dodoi{10.1093/mnras/sty1621}

\bibitem[{Eldridge {et~al.}(2008)Eldridge, Izzard, \& Tout}]{Eldridge2008}
Eldridge, J.~J., Izzard, R.~G., \& Tout, C.~A. 2008, \mnras, 384, 1109,
  \dodoi{10.1111/j.1365-2966.2007.12738.x}

\bibitem[{Ferraro {et~al.}(2006)Ferraro, Sabbi, Gratton, Piotto, Lanzoni,
  Carretta, Rood, Sills, Pecci, Moehler, Beccari, Lucatello, \&
  Compagni}]{Ferraro2006}
Ferraro, F.~R., Sabbi, E., Gratton, R., {et~al.} 2006, \apj, 647, L53,
  \dodoi{10.1086/507327}

\bibitem[{{Ferraro} {et~al.}(2009){Ferraro}, {Beccari}, {Dalessandro},
  {Lanzoni}, {Sills}, {Rood}, {Pecci}, {Karakas}, {Miocchi}, \&
  {Bovinelli}}]{Ferraro2009}
{Ferraro}, F.~R., {Beccari}, G., {Dalessandro}, E., {et~al.} 2009, \nat, 462,
  1028, \dodoi{10.1038/nature08607}

\bibitem[{Ferraro {et~al.}(2018)Ferraro, Lanzoni, Raso, Nardiello, Dalessandro,
  Vesperini, Piotto, Pallanca, Beccari, Bellini, Libralato, Anderson, Aparicio,
  Bedin, Cassisi, Milone, Ortolani, Renzini, Salaris, \& van~der
  Marel}]{Ferraro2018}
Ferraro, F.~R., Lanzoni, B., Raso, S., {et~al.} 2018, \apj, 860, 36,
  \dodoi{10.3847/1538-4357/aac01c}

\bibitem[{{Ferraro} {et~al.}(2023){Ferraro}, {Mucciarelli}, {Lanzoni},
  {Pallanca}, {Cadelano}, {Billi}, {Sills}, {Vesperini}, {Dalessandro},
  {Beccari}, {Monaco}, \& {Mateo}}]{Ferraro2023}
{Ferraro}, F.~R., {Mucciarelli}, A., {Lanzoni}, B., {et~al.} 2023, NatCo, 14,
  2584, \dodoi{10.1038/s41467-023-38153-w}

\bibitem[{Fleck {et~al.}(2006)Fleck, Boily, Lançon, \& Deiters}]{Fleck2006}
Fleck, J.-J., Boily, C.~M., Lançon, A., \& Deiters, S. 2006, \mnras, 369,
  1392, \dodoi{10.1111/j.1365-2966.2006.10390.x}

\bibitem[{{Fregeau} {et~al.}(2004){Fregeau}, {Cheung}, {Portegies Zwart}, \&
  {Rasio}}]{Fregeau2004}
{Fregeau}, J.~M., {Cheung}, P., {Portegies Zwart}, S.~F., \& {Rasio}, F.~A.
  2004, \mnras, 352, 1, \dodoi{10.1111/j.1365-2966.2004.07914.x}

\bibitem[{Gosnell {et~al.}(2014)Gosnell, Mathieu, Geller, Sills, Leigh, \&
  Knigge}]{Gosnell2014}
Gosnell, N.~M., Mathieu, R.~D., Geller, A.~M., {et~al.} 2014, \apjl, 783, L8,
  \dodoi{10.1088/2041-8205/783/1/L8}

\bibitem[{Han {et~al.}(1994)Han, Podsiadlowski, \& Eggleton}]{Han1994}
Han, Z., Podsiadlowski, P., \& Eggleton, P.~P. 1994, \mnras, 270, 121,
  \dodoi{10.1093/mnras/270.1.121}

\bibitem[{Han {et~al.}(1995)Han, Podsiadlowski, \& Eggleton}]{Han1995}
---. 1995, \mnras, 272, 800, \dodoi{10.1093/mnras/272.4.800}

\bibitem[{{Hills} \& {Day}(1976)}]{Hills1976}
{Hills}, J.~G., \& {Day}, C.~A. 1976, \aplett, 17, 87

\bibitem[{Hurley {et~al.}(2000)Hurley, Pols, \& Tout}]{Hurley2000}
Hurley, J.~R., Pols, O.~R., \& Tout, C.~A. 2000, \mnras, 315, 543,
  \dodoi{10.1046/j.1365-8711.2000.03426.x}

\bibitem[{Hypki \& Giersz(2013)}]{Hypki2013}
Hypki, A., \& Giersz, M. 2013, \mnras, 429, 1221, \dodoi{10.1093/mnras/sts415}

\bibitem[{Jiang(2022)}]{Jiang2022}
Jiang, D. 2022, \apj, 940, 97, \dodoi{10.3847/1538-4357/ac9a42}

\bibitem[{Jiang {et~al.}(2017)Jiang, Chen, Li, \& Han}]{Jiang2017}
Jiang, D., Chen, X., Li, L., \& Han, Z. 2017, \apj, 849, 100,
  \dodoi{10.3847/1538-4357/aa8ee1}

\bibitem[{Jiang {et~al.}(2013)Jiang, Han, \& Li}]{Jiang2013}
Jiang, D., Han, Z., \& Li, L. 2013, \mnras, 438, 859,
  \dodoi{10.1093/mnras/stt2252}

\bibitem[{{King}(1962)}]{King1962}
{King}, I. 1962, \aj, 67, 471, \dodoi{10.1086/108756}

\bibitem[{{Kippenhahn} \& {Weigert}(1967)}]{Kippenhahn1967}
{Kippenhahn}, R., \& {Weigert}, A. 1967, \zap, 65, 251

\bibitem[{Knigge {et~al.}(2009)Knigge, Leigh, \& Sills}]{Knigge2009}
Knigge, C., Leigh, N., \& Sills, A. 2009, \nat, 457, 288,
  \dodoi{10.1038/nature07635}

\bibitem[{Leigh {et~al.}(2007)Leigh, Sills, \& Knigge}]{Leigh2007}
Leigh, N., Sills, A., \& Knigge, C. 2007, \apj, 661, 210,
  \dodoi{10.1086/514330}

\bibitem[{{Leiner} {et~al.}(2019){Leiner}, {Mathieu}, {Vanderburg}, {Gosnell},
  \& {Smith}}]{Leiner2019}
{Leiner}, E., {Mathieu}, R.~D., {Vanderburg}, A., {Gosnell}, N.~M., \& {Smith},
  J.~C. 2019, \apj, 881, 47, \dodoi{10.3847/1538-4357/ab2bf8}

\bibitem[{Li {et~al.}(2018)Li, Deng, Grijs, Jiang, \& Xin}]{Li2018}
Li, C., Deng, L., Grijs, R.~d., Jiang, D., \& Xin, Y. 2018, \apj, 856, 25,
  \dodoi{10.3847/1538-4357/aaad65}

\bibitem[{Lovisi {et~al.}(2010)Lovisi, Mucciarelli, Ferraro, Lucatello,
  Lanzoni, Dalessandro, Beccari, Rood, Sills, Pecci, Gratton, \&
  Piotto}]{Lovisi2010}
Lovisi, L., Mucciarelli, A., Ferraro, F.~R., {et~al.} 2010, \apjl, 719, L121,
  \dodoi{10.1088/2041-8205/719/2/L121}

\bibitem[{Lu {et~al.}(2010)Lu, Deng, \& Zhang}]{Lu2010}
Lu, P., Deng, L.~C., \& Zhang, X.~B. 2010, \mnras, 409, 1013,
  \dodoi{10.1111/j.1365-2966.2010.17356.x}

\bibitem[{Mapelli {et~al.}(2007)Mapelli, Ripamonti, Tolstoy, Sigurdsson, Irwin,
  \& Battaglia}]{Mapelli2007}
Mapelli, M., Ripamonti, E., Tolstoy, E., {et~al.} 2007, \mnras, 380, 1127,
  \dodoi{10.1111/j.1365-2966.2007.12148.x}

\bibitem[{Mathieu \& Geller(2009)}]{Mathieu2009}
Mathieu, R.~D., \& Geller, A.~M. 2009, \nat, 462, 1032,
  \dodoi{10.1038/nature08568}

\bibitem[{McCrea(1964)}]{McCrea1964}
McCrea, W.~H. 1964, \mnras, 128, 147, \dodoi{10.1093/mnras/128.2.147}

\bibitem[{McLaughlin \& Marel(2005)}]{Mclaughlin2005}
McLaughlin, D.~E., \& Marel, R. P. v.~d. 2005, \apjs, 161, 304,
  \dodoi{10.1086/497429}

\bibitem[{{Miller} \& {Scalo}(1979)}]{Miller1979}
{Miller}, G.~E., \& {Scalo}, J.~M. 1979, \apjs, 41, 513, \dodoi{10.1086/190629}

\bibitem[{Milliman {et~al.}(2015)Milliman, Mathieu, \& Schuler}]{Milliman2015}
Milliman, K.~E., Mathieu, R.~D., \& Schuler, S.~C. 2015, \apj, 150, 84,
  \dodoi{10.1088/0004-6256/150/3/84}

\bibitem[{Milone {et~al.}(2012)Milone, Piotto, Bedin, Aparicio, Anderson,
  Sarajedini, Marino, Moretti, Davies, Chaboyer, Dotter, Hempel, Marín-Franch,
  Majewski, Paust, Reid, Rosenberg, \& Siegel}]{Milone2012}
Milone, A.~P., Piotto, G., Bedin, L.~R., {et~al.} 2012, \aap, 540, A16,
  \dodoi{10.1051/0004-6361/201016384}

\bibitem[{Milone {et~al.}(2023)Milone, Cordoni, Marino, D'Antona, Bellini,
  Di~Criscienzo, Dondoglio, Lagioia, Langer, Legnardi, Libralato, Baumgardt,
  Bettinelli, Cavecchi, de~Grijs, Deng, Hastings, Li, \& {et al.}}]{Milone2023}
Milone, A.~P., Cordoni, G., Marino, A.~F., {et~al.} 2023, \aap,
  \dodoi{10.1051/0004-6361/202244798}

\bibitem[{{Mohandasan} {et~al.}(2024){Mohandasan}, {Milone}, {Cordoni},
  {Dondoglio}, {Lagioia}, {Legnardi}, {Ziliotto}, {Jang}, {Marino}, \&
  {Carlos}}]{Mohandasan2024}
{Mohandasan}, A., {Milone}, A.~P., {Cordoni}, G., {et~al.} 2024, \aap, 681,
  A42, \dodoi{10.1051/0004-6361/202347424}

\bibitem[{Nelson \& Eggleton(2001)}]{Nelson2001}
Nelson, C.~A., \& Eggleton, P.~P. 2001, \apj, 552, 664, \dodoi{10.1086/320560}

\bibitem[{Nine {et~al.}(2023)Nine, Mathieu, Gosnell, \& Leiner}]{Nine2023}
Nine, A.~C., Mathieu, R.~D., Gosnell, N.~M., \& Leiner, E.~M. 2023, \apj, 944,
  145, \dodoi{10.3847/1538-4357/acb046}

\bibitem[{{O'Donnell}(1994)}]{Odon1994}
{O'Donnell}, J.~E. 1994, \apj, 422, 158, \dodoi{10.1086/173713}

\bibitem[{Panthi {et~al.}(2023)Panthi, Vaidya, Vernekar, Subramaniam, Jadhav,
  \& Agarwal}]{Panthi2023}
Panthi, A., Vaidya, K., Vernekar, N., {et~al.} 2023, \mnras, 527, 8325,
  \dodoi{10.1093/mnras/stad3750}

\bibitem[{{Perets} \& {Fabrycky}(2009)}]{Perets2009}
{Perets}, H.~B., \& {Fabrycky}, D.~C. 2009, \apj, 697, 1048,
  \dodoi{10.1088/0004-637X/697/2/1048}

\bibitem[{Perets \& Kratter(2012)}]{perets2012}
Perets, H.~B., \& Kratter, K.~M. 2012, \apj, 760, 99,
  \dodoi{10.1088/0004-637X/760/2/99}

\bibitem[{Piotto {et~al.}(2004)Piotto, Angeli, King, Djorgovski, Bono, Cassisi,
  Meylan, Recio-Blanco, Rich, \& Davies}]{Piotto2004}
Piotto, G., Angeli, F.~D., King, I.~R., {et~al.} 2004, \apjl, 604, L109,
  \dodoi{10.1086/383617}

\bibitem[{Pols {et~al.}(1998)Pols, Schröder, Hurley, Tout, \&
  Eggleton}]{pols1998}
Pols, O.~R., Schröder, K.-P., Hurley, J.~R., Tout, C.~A., \& Eggleton, P.~P.
  1998, \mnras, 298, 525, \dodoi{10.1046/j.1365-8711.1998.01658.x}

\bibitem[{Pols {et~al.}(1995)Pols, Tout, Eggleton, \& Han}]{Pols1995}
Pols, O.~R., Tout, C.~A., Eggleton, P.~P., \& Han, Z. 1995, \mnras, 274, 964,
  \dodoi{10.1093/mnras/274.3.964}

\bibitem[{{Popham} \& {Narayan}(1991)}]{Popham1991}
{Popham}, R., \& {Narayan}, R. 1991, \apj, 370, 604, \dodoi{10.1086/169847}

\bibitem[{Rain {et~al.}(2024)Rain, Pera, Perren, Benvenuto, Panei, Vito,
  Carraro, \& Villanova}]{Rain2024}
Rain, M.~J., Pera, M.~S., Perren, G.~I., {et~al.} 2024, \aap, 685, A33,
  \dodoi{10.1051/0004-6361/202347499}

\bibitem[{Rao {et~al.}(2023)Rao, Bhattacharya, Vaidya, \& Agarwal}]{Rao2023}
Rao, K.~K., Bhattacharya, S., Vaidya, K., \& Agarwal, M. 2023, \mnras, 518, L7,
  \dodoi{10.1093/mnrasl/slac122}

\bibitem[{{Rao} {et~al.}(2022){Rao}, {Vaidya}, {Agarwal}, {Panthi}, {Jadhav},
  \& {Subramaniam}}]{Rao2022}
{Rao}, K.~K., {Vaidya}, K., {Agarwal}, M., {et~al.} 2022, \mnras, 516, 2444,
  \dodoi{10.1093/mnras/stac2241}

\bibitem[{{Raso} {et~al.}(2020){Raso}, {Libralato}, {Bellini}, {Ferraro},
  {Lanzoni}, {Cadelano}, {Pallanca}, {Dalessandro}, {Piotto}, {Anderson}, \&
  {Sohn}}]{Raso2020}
{Raso}, S., {Libralato}, M., {Bellini}, A., {et~al.} 2020, \apj, 895, 15,
  \dodoi{10.3847/1538-4357/ab8ae7}

\bibitem[{{Ren} {et~al.}(2023){Ren}, {Li}, {Ma}, {Cheng}, {Huang}, {Tang}, \&
  {Hu}}]{Ren2023}
{Ren}, L., {Li}, C., {Ma}, B., {et~al.} 2023, \apjs, 264, 39,
  \dodoi{10.3847/1538-4365/aca09e}

\bibitem[{Rucinski(2000)}]{Rucinski2000}
Rucinski, S.~M. 2000, \apj, 120, 319, \dodoi{10.1086/301417}

\bibitem[{{Sandage}(1953)}]{Sandage1953}
{Sandage}, A.~R. 1953, \aj, 58, 61, \dodoi{10.1086/106822}

\bibitem[{{Schirbel} {et~al.}(2015){Schirbel}, {Mel{\'e}ndez}, {Karakas},
  {Ram{\'\i}rez}, {Castro}, {Faria}, {Lugaro}, {Asplund}, {Tucci Maia}, {Yong},
  {Howes}, \& {do Nascimento}}]{Schirbel2015}
{Schirbel}, L., {Mel{\'e}ndez}, J., {Karakas}, A.~I., {et~al.} 2015, \aap, 584,
  A116, \dodoi{10.1051/0004-6361/201527303}

\bibitem[{Schneider {et~al.}(2019)Schneider, Ohlmann, Podsiadlowski, Röpke,
  Balbus, Pakmor, \& Springel}]{Schneider2019}
Schneider, F. R.~N., Ohlmann, S.~T., Podsiadlowski, P., {et~al.} 2019, \nat,
  574, 211, \dodoi{10.1038/s41586-019-1621-5}

\bibitem[{Simunovic {et~al.}(2014)Simunovic, Puzia, \& Sills}]{Simunovic2014}
Simunovic, M., Puzia, T.~H., \& Sills, A. 2014, \apjl, 795, L10,
  \dodoi{10.1088/2041-8205/795/1/L10}

\bibitem[{{Sollima} {et~al.}(2008){Sollima}, {Lanzoni}, {Beccari}, {Ferraro},
  \& {Fusi Pecci}}]{Sollima2008}
{Sollima}, A., {Lanzoni}, B., {Beccari}, G., {Ferraro}, F.~R., \& {Fusi Pecci},
  F. 2008, \aap, 481, 701, \dodoi{10.1051/0004-6361:20079082}

\bibitem[{{Sun} {et~al.}(2024){Sun}, {Levina}, {Gossage}, {Kalogera}, {Leiner},
  {Geller}, \& {Doctor}}]{Sun2024}
{Sun}, M., {Levina}, S., {Gossage}, S., {et~al.} 2024, \apj, 969, 8,
  \dodoi{10.3847/1538-4357/ad47c1}

\bibitem[{{Sun} \& {Mathieu}(2023)}]{Sun2023}
{Sun}, M., \& {Mathieu}, R.~D. 2023, \apj, 944, 89,
  \dodoi{10.3847/1538-4357/acacf7}

\bibitem[{Sun {et~al.}(2021)Sun, Mathieu, Leiner, \& Townsend}]{Sun2021}
Sun, M., Mathieu, R.~D., Leiner, E.~M., \& Townsend, R. H.~D. 2021, \apj, 908,
  7, \dodoi{10.3847/1538-4357/abd402}

\bibitem[{Sun {et~al.}(2018)Sun, Li, Grijs, \& Deng}]{Sun2018}
Sun, W., Li, C., Grijs, R.~D., \& Deng, L. 2018, \apj, 862, 133,
  \dodoi{10.3847/1538-4357/aacc6b}

\bibitem[{{Tillich} {et~al.}(2010){Tillich}, {Przybilla}, {Scholz}, \&
  {Heber}}]{Tillich2010}
{Tillich}, A., {Przybilla}, N., {Scholz}, R.~D., \& {Heber}, U. 2010, \aap,
  517, A36, \dodoi{10.1051/0004-6361/201014315}

\bibitem[{Wang {et~al.}(2022)Wang, Langer, Schootemeijer, Milone, Hastings, Xu,
  Bodensteiner, Sana, Castro, Lennon, Marchant, Koter, \& Mink}]{Wang2022}
Wang, C., Langer, N., Schootemeijer, A., {et~al.} 2022, \nat, 6, 480,
  \dodoi{10.1038/s41550-021-01597-5}

\bibitem[{Wang {et~al.}(2024)Wang, Deng, Pang, Wang, Grijs, Milone, \&
  Li}]{Wang2024}
Wang, L., Deng, L., Pang, X., {et~al.} 2024, \apj, 969, 21,
  \dodoi{10.3847/1538-4357/ad499c}

\bibitem[{{Wang} {et~al.}(2020){Wang}, {Iwasawa}, {Nitadori}, \&
  {Makino}}]{Wang2020}
{Wang}, L., {Iwasawa}, M., {Nitadori}, K., \& {Makino}, J. 2020, \mnras, 497,
  536, \dodoi{10.1093/mnras/staa1915}

\bibitem[{{Wang} {et~al.}(2022){Wang}, {Tanikawa}, \& {Fujii}}]{WangL2022}
{Wang}, L., {Tanikawa}, A., \& {Fujii}, M.~S. 2022, \mnras, 509, 4713,
  \dodoi{10.1093/mnras/stab3255}

\bibitem[{{Xin} {et~al.}(2015){Xin}, {Ferraro}, {Lu}, {Deng}, {Lanzoni},
  {Dalessandro}, \& {Beccari}}]{Xin2015}
{Xin}, Y., {Ferraro}, F.~R., {Lu}, P., {et~al.} 2015, \apj, 801, 67,
  \dodoi{10.1088/0004-637X/801/1/67}

\bibitem[{Yakut \& Eggleton(2005)}]{Yakut2005}
Yakut, K., \& Eggleton, P.~P. 2005, \apj, 629, 1055, \dodoi{10.1086/431300}

\end{thebibliography}
\bibliographystyle{aasjournal}

\end{CJK*}
\end{document}